\documentclass[prd,amsmath,amssymb,superscriptaddress,nofootinbib,twocolumn]{revtex4}

\usepackage[T1]{fontenc}
\usepackage[latin9]{inputenc}
\usepackage{xcolor}
\usepackage{bm}
\usepackage[normalem]{ulem}
\usepackage{babel}
\usepackage{textcomp}
\usepackage{amstext}
\usepackage{graphicx}
\usepackage{esint}

\usepackage[unicode=true,pdfusetitle,
 bookmarks=true,bookmarksnumbered=false,bookmarksopen=false,
 breaklinks=false,pdfborder={0 0 1},backref=false,colorlinks=true]
 {hyperref}
\hypersetup{
 pdfborderstyle=,linkcolor=blue,citecolor=cyan}

\newcommand{\nn}{\nonumber}

\newcommand{\beq}{\begin{equation}}
\newcommand{\eeq}{\end{equation}}
\newcommand{\bqa}{\begin{eqnarray}}
\newcommand{\eqa}{\end{eqnarray}}

\newcommand{\bseq}{\begin{subequations}}
\newcommand{\eseq}{\end{subequations}}

\makeatletter

\providecommand{\tabularnewline}{\\}

\usepackage{babel}
\usepackage{xcolor}

\begin{document}

\title{Azimuthal modulation in light-by-light scattering from ultraperipheral collisions at LHC
}

\author{Yu Jia~\footnote{jiay@ihep.ac.cn}}
\affiliation{Institute of High Energy Physics,
Chinese Academy of Sciences, Beijing 100049, China\vspace{0.2cm}}
\author{Shuo Lin~\footnote{linshuo@mail.ustc.edu.cn}}
\affiliation{Department of Modern Physics, University of Science and Technology
of China, Anhui 230026, China\vspace{0.2cm}}
\author{Jian Zhou~\footnote{jzhou@sdu.edu.cn}}
\affiliation{Key Laboratory of
Particle Physics and Particle Irradiation (MOE),Institute of
Frontier and Interdisciplinary Science, Shandong University,
(QingDao), Shandong 266237, China \vspace{0.2cm}}
\affiliation{Southern Center for Nuclear-Science Theory (SCNT), Institute of Modern Physics, Chinese Academy of Sciences, HuiZhou, Guangdong
516000, China\vspace{0.2cm}}
\author{Ya-jin Zhou~\footnote{zhouyj@sdu.edu.cn}}
\affiliation{Key Laboratory of
Particle Physics and Particle Irradiation (MOE),Institute of
Frontier and Interdisciplinary Science, Shandong University,
(QingDao), Shandong 266237, China \vspace{0.2cm}}

\date{\today}

\begin{abstract}
Elastic light-by-light (LbL) scattering, one of the most fascinating processes in the Standard Model (SM), has recently been observed in the ultraperipheral collisions (UPCs) of relativistic heavy ions in the {\tt Atlas} and {\tt CMS} experiments at the Large Hadron Collider ({\tt LHC}).
Recognizing that the incident quasi-real photons in LbL scattering are strongly linearly polarized, we re-investigate the LbL scattering in UPCs  by incorporating the joint dependence of the impact parameter and transverse momenta of the incident photons. 
We show that the linear polarization of incident photons generates a sizable $\cos2\phi$-type azimuthal modulation,
which awaits the test in  future {\tt LHC} and {\tt EIC} experiments.
\end{abstract}

\maketitle

\noindent{\color{blue}\it Introduction---} In classical Maxwell theory, electromagnetic waves in  vacuum obey the principle of linear superposition.  However, this simple law is profoundly changed at quantum mechanical level,
since the photon can acquire self-interaction via quantum fluctuating into virtual charged particles.
As the direct manifestation of the vacuum polarization and quantum nonlinearity,
the elastic light-by-light (LbL) scattering, $\gamma\gamma\to\gamma\gamma$,
is widely viewed as one of the  most fascinating and fundamental processes in Standard Model (SM)~\cite{Dunne:2012vv}.

Historically, Halpern~\cite{Halpern:1933dya} was the first to  qualitatively study
the LbL scattering about nine decades ago.
Within the framework of Dirac's hole theory~\cite{Dirac:1930ek}, {\it viz.},
the precursor of quantum electrodynamics (QED), during the period of 1935-36 Heisenberg, Euler and Kockel~\cite{Euler:1935zz,Euler:1935qgl,Heisenberg:1936nmg}
systematically investigated the LbL scattering in the low frequency limit ($\omega\ll m_e$).
Shortly after, Akhieser, Landau and Pomeranchuk~\cite{Akhiezer:1936vzu} also studied the LbL scattering in the high frequency limit.
With the aid of the Feynman diagram technique,  in 1951 Karplus and Neuman~\cite{Karplus:1950zz} presented a thorough analysis
for the polarized cross sections of $\gamma\gamma\to\gamma\gamma$, valid with an arbitrary photon frequency.
As a loop induced process, the high-energy LbL scattering nowadays serves a fertile playground to search for
the footprints of various beyond SM scenarios, such as axion-like particle~\cite{Knapen:2016moh},
magnetic monopole~\cite{Ginzburg:1998vb}, vector-like fermion~\cite{Fichet:2014uka},
 extra dimensions~\cite{Atag:2010bh}, Born-Infeld extension of QED~\cite{Ellis:2017edi},
and spacetime noncommutativity~\cite{Horvat:2020ycy}, {\it etc.}

The {\it indirect} evidences of LbL scattering have long been accumulated from various sources,
including the anomalous magnetic moments of the muon~\cite{Blum:2014oka},
and a flurry of strong field related phenomena
exemplified by Delbr\"{u}ck scattering~\citep{Delbruck:1933pla,Jarlskog:1973aui}, photon splitting~\citep{Adler:1971wn,Adler:1970gg}, and
vacuum birefringence~\citep{PVLAS:2008iru,Baier:1967zzc,Bialynicka-Birula:1970nlh,Toll:1952rq}.
On the other hand, since the production rate of $\gamma\gamma\to\gamma\gamma$ is suppressed by a factor of $\alpha^4$,
it turns out to be rather challenging to {\it directly} observe the elusive on-shell LbL process in collider experiments.
The photon-photon collider using the laser Compton scattering has been proposed to
detect LbL in the MeV range~\cite{Homma:2015fva, Micieli:2016vpj, Takahashi:2018uut}.
While the copious background severely hinders the prospect of singling out the rare signal events
in the $e^+e^-$ machine~\cite{Beloborodov:2023ajj},
the recent search of the LbL scattering events in $pp$ collision in {\tt TOTEM} and {\tt CMS} detectors~\cite{TOTEM:2021zxa,TOTEM:2023ewz}
has also yielded negative results.

During the past two decades, ultraperipherad collisions (UPCs) in relativistic heavy ion experiments proves to be a fruitful playground for
the program of two-photon physics, thanks to the exceptionally high quasi-real photon flux of the ion beam brought by the $Z^4$ enhancement ($Z$ denotes the nuclear charge number)~\cite{Baltz:2007kq}.
Following the suggestion by d'Enterria and da Silveira in 2013~\cite{dEnterria:2013zqi},
by analyzing the $2.2\;{\rm nb}^{-1}$ UPC data samples,  {\tt Atlas}~\citep{ATLAS:2017fur,ATLAS:2019azn,ATLAS:2020hii}
and {\tt CMS}~\citep{CMS:2018erd} experiments at {\tt LHC} have recently
reported the first direct observation of the long-sought LbL scattering process, in the range of $5<Q<30$ GeV ($Q$ denotes the invariant mass of
the outgoing photon pair). The measured invariant mass distribution generally  agrees with the  SM predictions~\cite{Klusek-Gawenda:2016euz, Harland-Lang:2018iur, Harland-Lang:2020veo,Shao:2022cly}
within experimental uncertainties, though noticeable tension remains between the data and predictions at low-$Q$.

In most recent studies of LbL scattering in UPCs, theoretical predictions rely on the standard collinear factorization approach, where the photon parton distribution function (PDF) of the heavy ion is typically computed using the Weizs\"{a}cker-Williams (WW) method~\citep{Fermi:1924tc,Williams:1934ad,vonWeizsacker:1934nji}, with the exception of the prediction made by {\tt SuperChic}~\cite{Harland-Lang:2020veo}. In the energy range $Q>5$ GeV, the $\gamma\gamma\to \gamma\gamma$ at the lowest-order in SM
is dictated by the QED box diagrams with leptons and quarks circling around in the loop.
In 2001 Bern {\it et al.}~\cite{Bern:2001dg} have computed the two-loop QCD and QED corrections in the ultrarelativistic limit,
with all the fermion masses neglected. Very recently, Ajjath {\it et al.}~\cite{AH:2023ewe} have
improved the result of \cite{Bern:2001dg}, {\it viz.},
by presenting the two-loop LbL amplitudes in the closed form with the mass for heavy quarks and $W$ bosons retained.
The effect of two-loop corrections is found to be rather modest.
Even after including the two-loop corrections, the predictions based on collinear factorization still notably undershoots
the {\tt Atlas} measurement for LbL scattering in Pb-Pb UPCs~\cite{AH:2023kor}.

It is worth noting that, the two-photon physics program in UPCs in the past few years has received a strong boost also by
another fundamental QED process, {\it viz.}, the Breit-Wheeler (BW) process $\gamma \gamma \rightarrow e^+ e^-$.
Recognizing that the photons coherently radiated off the relativistic heavy ions are linearly polarized, in 2019 Li, Zhou and Zhou~\cite{Li:2019sin,Li:2019yzy}
predicted a substantial $\cos 4\phi $-type dilepton azimuthal asymmetry in the BW process,
which was soon confirmed by  gold-gold UPC in {\tt STAR} experiment~\cite{STAR:2019wlg}.
Since then, a surge of theoretical endeavors have been devoted to leveraging
the linear polarization of coherent photons to probe the nuclear structure~\cite{Hagiwara:2020juc,Hagiwara:2021xkf,Xing:2020hwh,Mantysaari:2023prg,Mantysaari:2022sux,Brandenburg:2024ksp,Lin:2024mnj,Shao:2024nor,Hagiwara:2017fye,Iancu:2023lel},
quarkonium and exotic hadron states~\cite{Niu:2022cug,Jiang:2024vuq}, and polarization-dependent new physics observables in UPCs~\cite{Xu:2022qme,Shao:2023bga},
alongside other traditional topics studied in UPCs~\cite{Klein:2016yzr,Bertulani:1987tz,Bertulani:2005ru,Baltz:2007kq,Zhao:2022dac,Copinger:2018ftr,Klein:2018fmp,
Klein:2020jom,Shao:2023zge,Shao:2022stc,Hu:2024bsm,Zhang:2024mql,Hencken:2004td,Zha:2018tlq,Brandenburg:2021lnj,Xiao:2020ddm,Wang:2021kxm,Wang:2022gkd,Lin:2022flv}.
On the experimental side, substantial advancements have been achieved in measuring UPC observables as well~\cite{ALICE:2019tqa,ALICE:2020ugp,ALICE:2021gpt,STAR:2022wfe,STAR:2019wlg,ALICE:2013wjo,CMS:2020skx,ATLAS:2020epq}.

The linearly-polarized photons radiated off the lead/gold ions carry typical transverse momentum about $1/R_A\sim 30$ MeV.
 Similar to the outgoing $e^-e^+$ pair in the BW process, the pair of scattered photons
always fly {\it nearly} back-to-back in the transverse plane, with the pair's transverse momentum, $q_\perp \sim 30$ MeV, being much smaller than that of each outgoing photon.
In this case, it is advantageous to retain the transverse momenta of the incident photons and employ the
transverse-momentum-dependent (TMD) factorization approach.
This work aims to reinvestigate the LbL scattering in UPCs, by incorporating the dependencies on
both the impact parameter $\bm{b}_\perp$ (the transverse distance between the centers of two colliding nuclei)
and $\bm{q}_\perp$ within the QED TMD factorization framework~\footnote{ The calculation of the joint $\bm{b}_\perp$ and $\bm{q}_\perp$ dependent cross section was first formulated in Refs.~\cite{Vidovic:1992ik} in the context of di-lepton production in UPCs.
This recipe  proves indispensable  for accurately accounting for the low transverse momentum spectrum of $e^-e^+$ pairs produced in UPCs at {\tt RHIC} and {\tt LHC}~\cite{Zha:2018tlq,Li:2019yzy,Brandenburg:2021lnj,Xiao:2020ddm,Wang:2021kxm,Wang:2022gkd,Lin:2022flv,Shi:2024gex}. 
}. We  compute the azimuthal-angle-dependent
differential cross section of LbL scattering, predicting a significant $\cos 2\phi$ azimuthal modulation induced by the linear polarization of the
coherent incident photons. This prediction eagerly awaits experimental verification in future measurements at the LHC.

\begin{figure}[hbp]
\centering\includegraphics[scale=0.33]{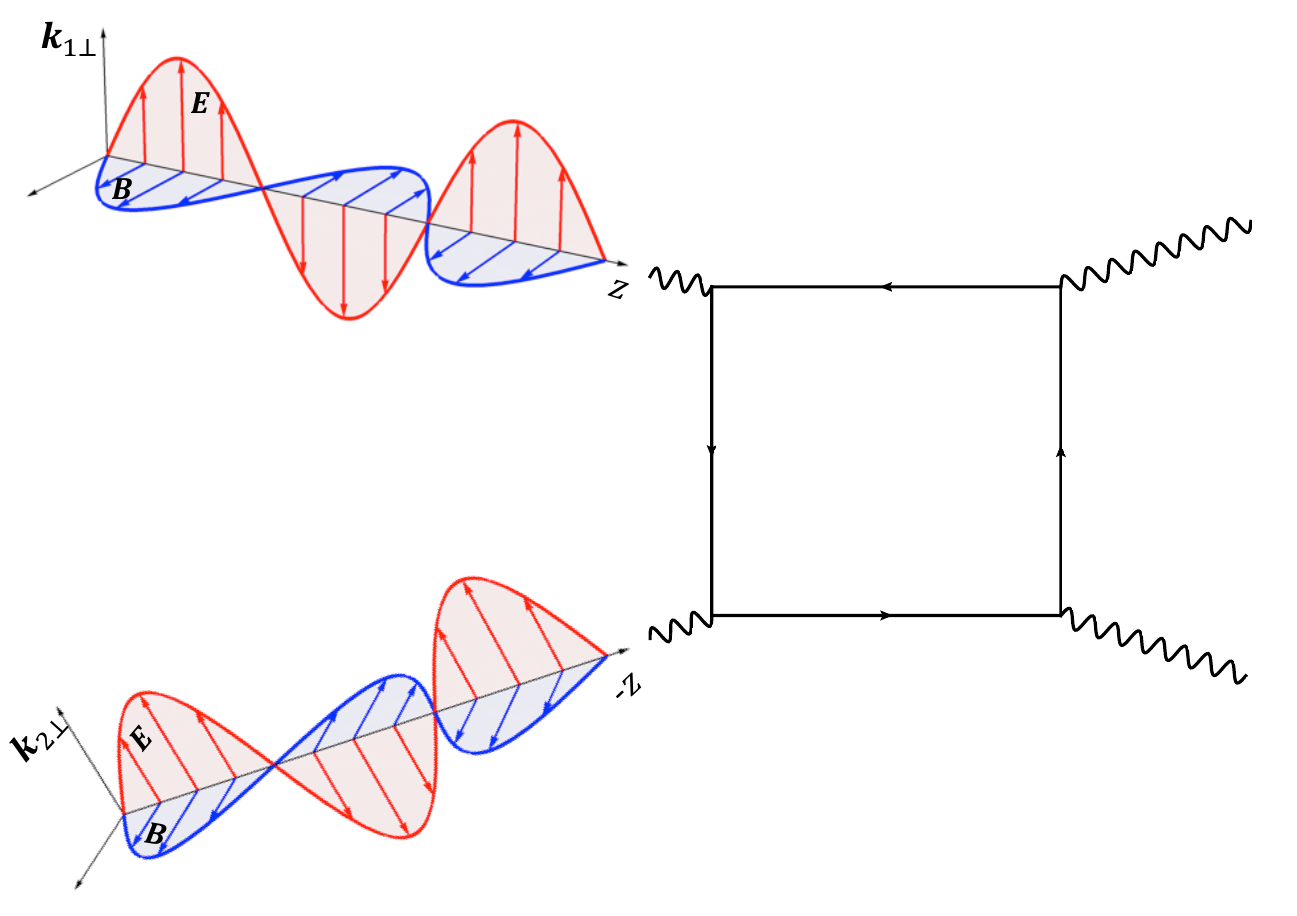}
\caption{A schematic representation of the lowest-order diagram for $\gamma\gamma \to \gamma\gamma$ in UPCs, highlighting the linear polarization of the photons in the initial state. \label{Fig:Cartoon:LbL}}
\end{figure}

\vspace{0.2cm}

\noindent{\color{blue}\it Azimuthal-dependent cross section of LbL scattering in UPC---}
Let us specialize to the reaction ${\rm Pb} + {\rm Pb}\to {\rm Pb}+{\rm Pb}+\gamma\gamma$.
A cartoon of the partonic LbL scattering is illustrated in Fig.~\ref{Fig:Cartoon:LbL}, with
the kinematics specified with
\beq
\gamma(x_{1}P+k_{1\perp})+\gamma(x_{2}\overline{P}+k_{1\perp})  \rightarrow \gamma(p_{1})+\gamma(p_{2}),
\eeq
where $x_{1}P$ and $x_{2}\overline{P}$ represent the light-cone momenta of the two incoming photons, their transverse momenta are denoted by
$\bm{k}_{1\perp}$ and $\bm{k}_{2\perp}$, and their four-momenta are given by $k_{1\perp} = (0, 0, \bm{k}_{1\perp})$ and $k_{2\perp} = (0, 0, \bm{k}_{2\perp})$. We emphasize  that the polarization vectors of the linearly-polarized quasi-real photons,
which are emitted from the relativistic lead ions, are parallel with their transverse momenta $\bm{k}_{1,2\bot}$.

It is convenient to introduce two transverse momenta ${\bm P}_\perp \equiv {{\bm p}_{1\perp}-{\bm p}_{2\perp}\over 2}$, and
${\bm q}_\perp\equiv {\bm p}_{1\perp}+{\bm p}_{2\perp}={\bm k}_{1\perp}+{\bm k}_{2\perp}$.
The azimuthal angle is defined by $\cos\phi \equiv \hat{\bm P}_\perp \cdot \hat{\bm q}_\perp$.
In the correlation limit where $q_\perp=\vert {\bm q}_\perp\vert \ll P_\perp =\vert {\bm P}_\perp\vert$,
one can approximate ${\bm P}_\perp\approx {\bm p}_{1\perp} \approx -{\bm p}_{2\bot}$.

 To incorporate both the $\bm{b}_\perp$ and $\bm{q}_\perp$ dependencies simultaneously,
one can derive the differential cross section for LbL scattering, following the formalism originally developed in
Refs.~\citep{Vidovic:1992ik,Hencken:1994my}, which is valid in the correlation limit.
After some manipulation, we cast the azimuthal-dependent differential cross section in the following convolutional form:
\bqa
& & \frac{d\sigma}{d^{2}\bm{p}_{1\perp}d^{2}\bm{p}_{2\perp}dy_{1}dy_{2}d^{2}\bm{b}_{\perp}}
\nn \\
 & = & \frac{1}{32\pi^{2}Q^{4}}\int d^{2}{\bm k}_{1\perp}d^{2}{\bm k}_{2\perp}\frac{d^{2}{\bm k}_{1\perp}'}{(2\pi)^{2}}
 \delta^{2}({\bm q}_{\perp}-{\bm k}_{1\perp}-{\bm k}_{2\perp})
 \nn \\
 & \times & e^{i({\bm k}_{1\perp}-{\bm k}_{1\perp}')\cdot {\bm b}_{\perp}}\Big\{ \cos(\phi_{1}-\phi_{2})\cos(\phi_{1}'-\phi_{2}')|M_{++}|^{2}
 \nn \\
 & + & \cos(\phi_{1}+\phi_{2})\cos(\phi_{1}'+\phi_{2}')|M_{+-}|^{2}
\label{eq:cross section}
 \\
 & -& \cos(\phi_{1}+\phi_{2})\cos(\phi_{1}'-\phi_{2}')M_{++}M_{+-}^{*}
 \nn \\
 &- & \cos(\phi_{1}-\phi_{2})\cos(\phi_{1}'+\phi_{2}')M_{+-}M_{++}^{*} \Big\}
\nn\\
 & \times  & {\cal F}(x_{1},{\bm k}_{1\perp}^{2})\,{\cal F}^{*}(x_{1},{\bm k}_{1\perp}'^{2})\,{\cal F}(x_{2},{\bm k}_{2\perp}^{2})\,{\cal F}^{*}(x_{2},{\bm k}_{2\perp}'^{2}),
 \nn
\eqa
 where  $y_{1,2}$ denote the rapidities of the outgoing photons, connected with the incident photons'
longitudinal momentum fractions $x_{1,2}$ through $x_{1,2}=\sqrt{\frac{{\bm P}_{\perp}^{2}}{s_{\rm NN}}}(e^{\pm y_1}+e^{\pm y_2})$
 ($\sqrt{s_{_{\rm NN}}}$  denotes the center-of-mass energy of each nucleon pair from the colliding nuclei). 
$\phi_{1,2}$ signify the azimuthal angles between ${\bm k}_{1,2\bot}$ and ${\bm P}_{\bot}$.
${\bm k}'_{1,2\perp}$  denote the transverse momenta of the incoming photons in the conjugated LbL scattering amplitude, constrained by momentum conservation ${\bm k}'_{1\perp}+{\bm k}'_{2\perp} ={\bm k}_{1\perp}+{\bm k}_{2\perp}$.
Meanwhile, $\phi_{1,2}'$ denote the azimuthal angles between ${\bm k}_{1,2\bot}'$ and ${\bm P}_{\bot}$~\footnote{Note that the
occurrence of ${\bm k}'_{i\perp}$ is the consequence of introducing the impact parameter dependence in our calculation.
The dependence of $\bm{b}_{\perp}$ enters the cross section through the phase factor in \eqref{eq:cross section}.
By integrating over $\bm{b}_{\bot}$ in \eqref{eq:cross section}, the joint $\bm{b}_\perp$ and $\bm{q}_\perp$ dependent
cross section reduces to the result obtained in the standard TMD factorization (For example, see Ref.~\cite{Jia:2024xzx} for a different context,
where the lead ion is replaced by the pointlike electron and positron).}.

$M_{\lambda_{1},\lambda_{2}}$ in  Eq.~\eqref{eq:cross section} is related with the
helicity amplitude of $\gamma(x_1 P,\lambda_1)\gamma(x_2\overline{P},\lambda_{2})\to\gamma(p_{1})+\gamma(p_{2})$.
The occurrence of $M_{\lambda_{1},\lambda_{2}}M_{\lambda_{1}^{\prime},\lambda_{2}^{\prime}}^{*}$ represents the short-hand
notation for
\beq
M_{\lambda_{1},\lambda_{2}}M_{\lambda_{1}^{\prime},\lambda_{2}^{\prime}}^{*} \equiv
 \sum_{\lambda_3,\lambda_4} M_{\lambda_{1},\lambda_{2},\lambda_3,\lambda_4} M^*_{\lambda_{1}^{\prime},\lambda_{2}^{\prime},\lambda_3,\lambda_4}.
\eeq
Note that the transverse momenta of the incoming photons are set to zero in the helicity amplitudes.
The parity invariance has been utilized to condense Eq.~\eqref{eq:cross section}.

The nonperturbative distribution function ${\cal F}(x,\bm{k}_{i\perp}^2)$ in Eq.~\eqref{eq:cross section}
characterizes the probability amplitude of finding a photon carries the prescribed light-momentum fraction and
transverse momenta inside a heavy nuclei.
It is intimately related to the standard photon TMD PDFs of a heavy nuclei:
\bqa
& & \int\frac{dy^{-}d^{2}y_{\bot}}{P^{+}(2\pi)^{3}}e^{ik\cdot y}\left\langle A|F_{+}^{\mu}(0)F_{+}^{\nu}(y)|A \right\rangle |_{y^{+}=0}
\nn \\
& = &\frac{\delta_{\bot}^{\mu\nu}}{2}xf_1(x,{\bm k}_{\bot}^{2})+\left(\frac{k_{\bot}^{\mu}k_{\bot}^{\nu}}{{\bm k}_{\bot}^{2}}-\frac{\delta_{\bot}^{\mu\nu}}{2}\right) x h_1^\bot(x,{\bm k}_{\bot}^{2}),
\eqa
where $\delta_{\bot}^{\mu\nu}$ denotes the transverse metric tensor,
$f_1$ and $h_1^{\bot}$ refer to the unpolarized and linearly-polarized photon TMD distributions, respectively. At small $x$, the TMD PDFs $f_1$ and $h_{1}^{\bot}$ can be simply related to the square of ${\cal F}$~\cite{Li:2019sin,Li:2019yzy}:
\beq
\label{eq:photonTMD}
 x f_1(x,\bm{k}_{\perp}^2) = x h_1^\perp(x,\bm{k}_{\perp}^2) =  \left| {\cal F} (x, \bm{k}^2_\perp) \right|^2.
\eeq

It is important to notice that the last two terms in the curly bracket in Eq.~\eqref{eq:cross section}
entail the interference between different helicity amplitudes, which is a direct consequence of the linear polarization of
the incoming quasi-real photons.
After integrating over the transverse momenta of incoming photons, the
angular correlations between ${\bm k}_{i\bot}$, ${\bm k}_{i\bot}'$ and ${\bm P}_{\bot}$ in the last two terms in the curly bracket
are converted into the angular correlation between ${\bm q}_\perp$ and ${\bm P}_\perp$.
The first two terms in the curly bracket contribute to the azimuthal-averaged cross section,
which differ from the expression for the azimuthal-averaged cross section derived in Refs.~\cite{Klusek-Gawenda:2016euz,Harland-Lang:2020veo,Shao:2022cly,AH:2023kor}
due to the intriguing entanglement between the impact parameter and polarization vectors of the coherent photons.

\vspace{0.2cm}

\noindent{\color{blue}\it Phenomenology---}
In the numerical analysis, we take the one-loop expressions of the helicity amplitudes for $\gamma\gamma\to\gamma\gamma$ from \cite{Bardin:2009gq}.
We include the contributions from the three charged leptons $e$, $\mu$, $\tau$, as well as five flavors of quarks, $u$, $d$, $s$, $c$ and $b$~\footnote{With $Q>5$ GeV,
one enters the energy range where the asymptotic freedom of QCD can be trusted and one can safely work at the quark level.
To avoid double counting, we have refrained from further including the contributions of the charged hadrons ($\pi^\pm$, $\rho^\pm$, $p$, $\Delta$, \ldots) in the loop.}.
For simplicity, we neglect the contribution from $t$ and $W$ since they are too heavy to render a noticeable impact
in the kinematic window $5<Q<30$ GeV adopted by the {\tt Atlas} and {\tt CMS} experiments. We treat $e$, $\mu$, $u$, $d$, $s$ as massless and take
$m_\tau=1.777$ GeV. We use the $\overline{\rm MS}$ masses for the two heavy quarks:
$\overline{m}_c(m_c)=1.275$ GeV and $\overline{m}_b(m_b)=4.196$ GeV~\cite{ParticleDataGroup:2024cfk}.

At low transverse momentum, the distribution amplitude ${\cal F}(x,\bm{k}_{\perp}^2)$ can be conveniently
calculated with the aid of the WW method, which has become a common practice in the field of UPC:
\beq
{\cal F}(x,\bm{k}_{\perp}^2)= {Z \sqrt{\alpha}\over \pi} |{\bm k}_{\perp}| {F({\bm k}_{\perp}^{2}+x^{2}M_p^2)\over {\bm k}_{\perp}^{2}+x^{2}M_p^2},
\label{eq:F}
\eeq
where $F(p^2)$ denotes the nuclear charge form factor, and $M_p$ denotes the proton mass. 
Following the {\tt STARlight} Monte Carlo generator~\cite{Klein:2016yzr},
we take the Woods-Saxon parameterization of the nuclear form factor:
\beq
F(p^2)  =  {4\pi\rho^0\over A\,p^3 }\left[\sin(pR_{A})-pR_{A}\cos(pR_{A})\right] {1\over a^{2}p^{2}+1},
\eeq
with $a=0.7$ fm. $A$ signifies the number of nucleons in the nucleus, $R_A \approx 1.1A^{1/3}$ fm denotes
the nuclear radius, and $\rho^{0}= 3A/(4\pi R^{3}_{A}) $ represents the average nucleon number density
in the nucleus.

\begin{table*}
\caption{Confrontation of various theoretical predictions with the {\tt Atlas} measurement for the
total cross section of LbL scattering in Pb+Pb UPCs at $\sqrt{s_{_{\rm NN}}}=5.02$  TeV.
\label{tab:total cross section}}
\centering%
\begin{tabular}{c|c||c|c|c|c|c}
\hline
 & Measured~\cite{ATLAS:2020hii} & Ref.~\cite{Klusek-Gawenda:2016euz} &  {\tt SuperChic v3.0}~\cite{Harland-Lang:2018iur} & {\tt gamma-UPC}~\cite{Shao:2022cly} & Ajjath {\it et al.}~\cite{AH:2023kor} & This work\tabularnewline
\hline
$\sigma(\mathrm{nb})$ & {\small{}120\textpm 17 (stat.)\textpm 13 (syst.)\textpm 4 (lumi.)}&$80\pm8$   & $78\pm8$ & $76$ & $81.2_{-0.9}^{+1.6}$ & {\small{}$83.6 \pm 3.1$}\tabularnewline
\hline
\end{tabular}

\end{table*}

\begin{figure}
\centering\includegraphics[scale=0.275]{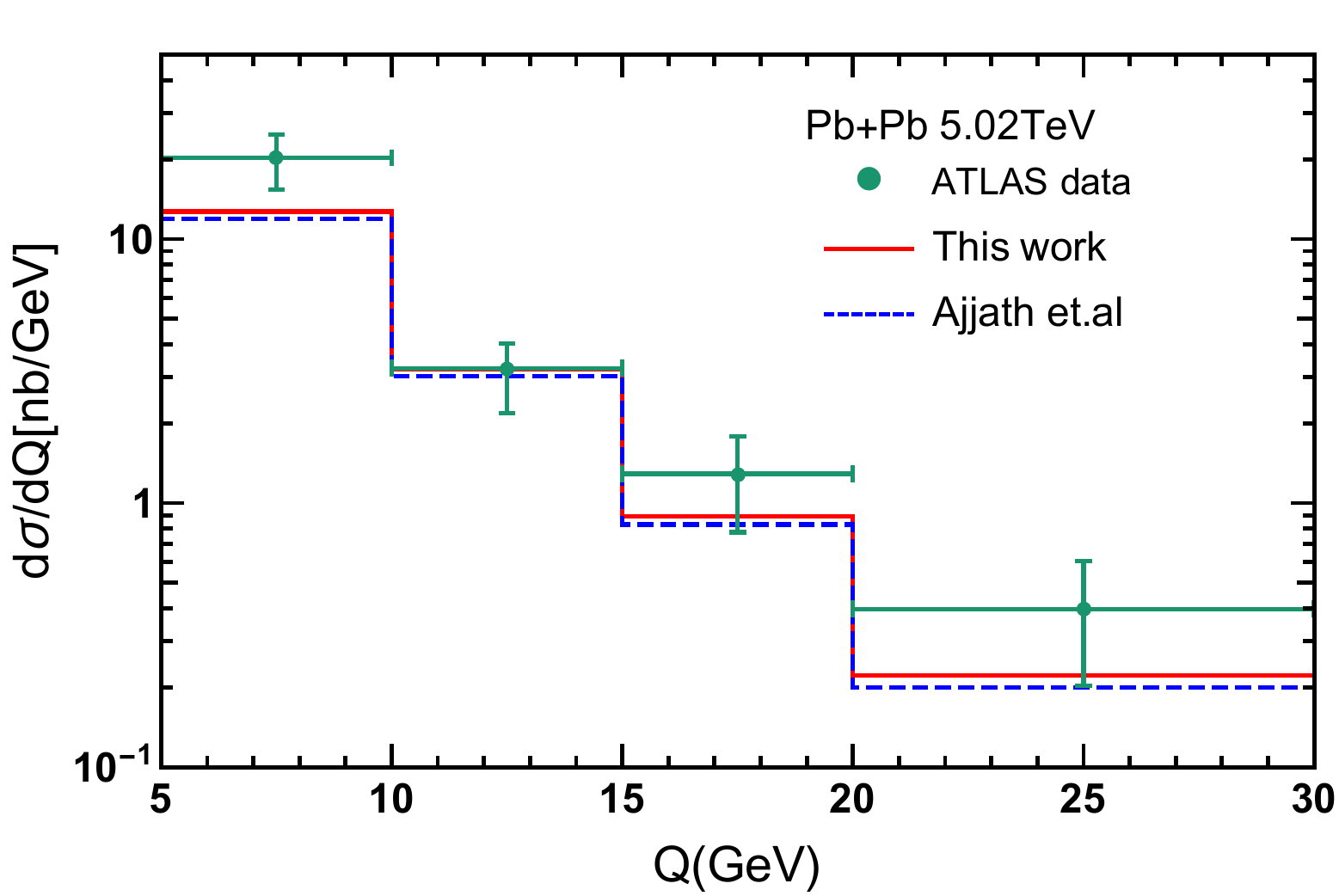}

\centering\includegraphics[scale=0.29]{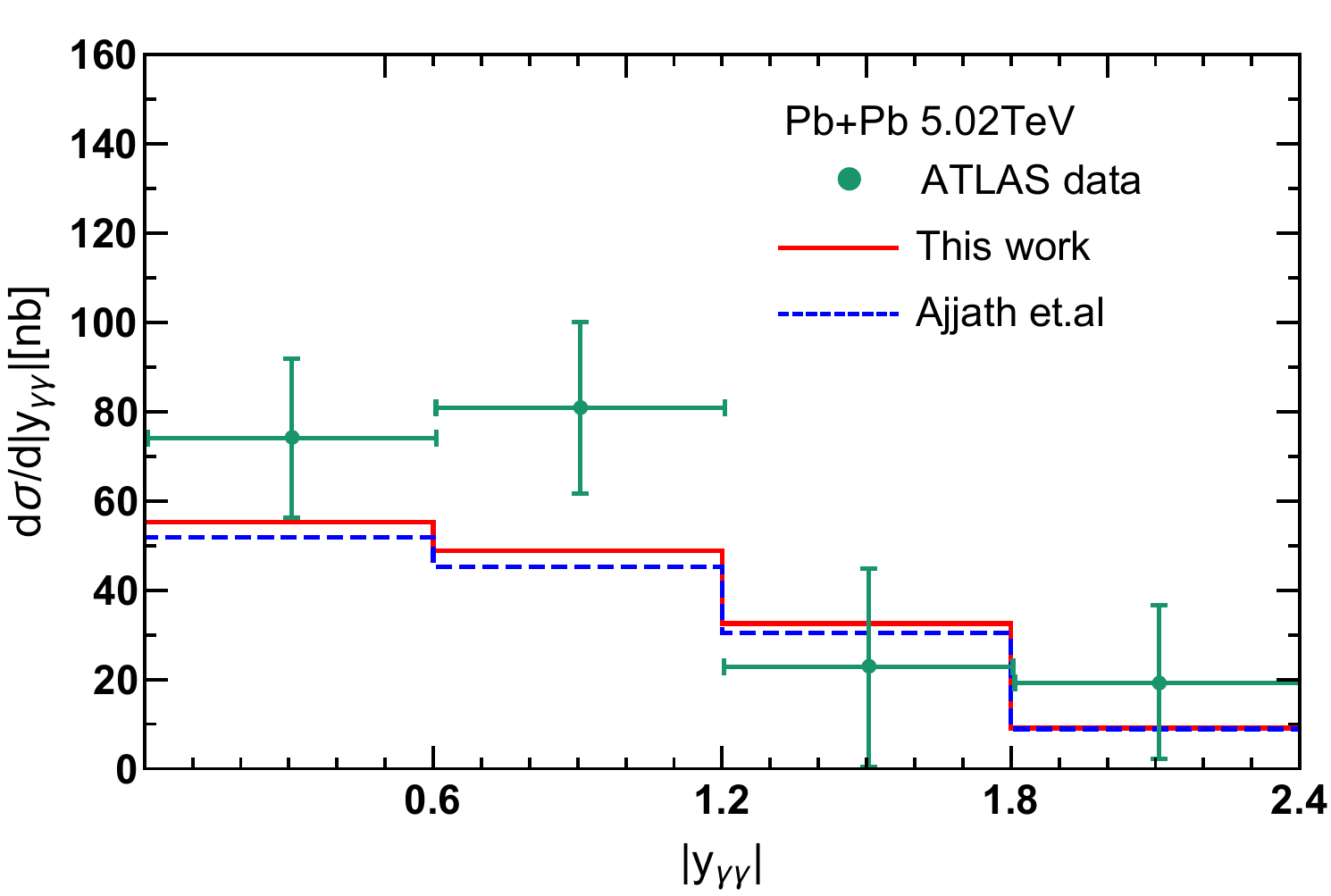}
\caption{The cross sections of LbL scattering  differential in 
invariant mass $Q$ (upper panel) and rapidity of the outgoing photon pair $y_{\gamma\gamma}$ (lower panel) in Pb-Pb UPCs at $\sqrt{s_{_{\rm NN}}}=5.02$ TeV.
The red solid line represents our TMD-based prediction re-scaled by the corresponding $K$ factor at each bin read off from Ref.~\cite{AH:2023kor}),
while the blue dashed line represents the prediction by Ajjath {\it et.al.}~\cite{AH:2023kor}.
\label{fig:Pt}}.
\end{figure}

Let us specialize to the Pb-Pb UPCs at $\sqrt{s_{_{\rm NN}}}=5.02$  TeV.
To comply with the kinematic cuts imposed by {\tt Atlas}~\cite{ATLAS:2020hii},
We require $Q > 5$ GeV and $p_{1,2\perp} > 2.5$ GeV, and we restrict the rapidity of each outgoing photon to the interval $|y_{1,2}| < 2.37$. Additionally, we constrain the total transverse momentum of the diphoton to be less than 1 GeV~\footnote{We find that the bulk of the predicted cross section
actually comes from the range $0< q_\bot < 200$ MeV, which is compatible with the correlation limit,
thereby justifying the application of TMD factorization.}.
Furthermore, we require that the acoplanarity, defined as $A_{\phi} = 1 - |\Delta\phi_{\gamma\gamma}|/\pi$, be smaller than 0.01, where $\Delta\phi_{\gamma\gamma}$ denotes the azimuthal angle between ${\bm p}_{1\perp}$ and ${\bm p}_{2\perp}$.
For the unrestricted UPC case, the impact parameter $ b_\perp $ is integrated from $2R_A$ to $\infty$.
The multi-dimensional integration in (\ref{eq:cross section}) is conducted numerically
by employing the package {\tt ZMCintegral}~\citep{Wu:2019tsf,Zhang:2019nhd}.

In Table~\ref{tab:total cross section} we juxtapose our numerical prediction for the total cross section of LbL scattering with the {\tt Atlas} measurement~\cite{ATLAS:2020hii},
along with the preceding theoretical predictions~\cite{Klusek-Gawenda:2016euz,Harland-Lang:2018iur,Shao:2022cly,AH:2023kor}.
From Table~\ref{tab:total cross section} one sees that all the previous studies appear to underestimate the measured value  to varying degrees.
Specifically, {\tt SuperChic v3.0}~\cite{Harland-Lang:2020veo} and {\tt gamma-UPC}~\cite{Shao:2022cly} utilize the one-loop LbL scattering amplitude,
while the two-loop QCD and QED corrections have been included in the recent work by Ajjath {\it et al.}~\citep{AH:2023kor}.
Apart from the standard one-loop LbL amplitude, Klusek-Gawenda, Lebiedowicz and Szczurek~\cite{Klusek-Gawenda:2016euz} have
also incorporated the vector-meson dominance and Regge contributions in their analysis.  Our numerical analysis finds that the total LbL scattering cross section for Pb+Pb UPC is $78.5 \pm 2.9$ nb. The uncertainty arises from several sources, including the intrinsic inaccuracies of Monte Carlo integration, uncertainties in the quark mass values. Specifically, we vary the strange quark mass from 0 to 0.1 GeV, the charm quark mass from 1.275 to 1.5 GeV, and the bottom quark mass from 4.196 to 4.75 GeV. It is worth noting that our predicted cross section of $78.5 \pm 2.9$ nb is slightly larger than the predictions from {\tt SuperChic v3.0}\cite{Harland-Lang:2020veo} and {\tt gamma-UPC}\cite{Shao:2022cly}, which also use the one-loop LbL amplitude as input.  Incorporating two-loop QCD/QED corrections\cite{AH:2023ewe} refines our prediction. Using a $K$ factor of 1.065 from \cite{AH:2023kor}, we adjust our result to $\sigma = 83.6 \pm 3.1
 nb$, as shown in Table~\ref{tab:total cross section}.

In Fig.~\ref{fig:Pt}, we plot the distributions of  the invariant mass and
rapidity of the outgoing photon pair in Pb-Pb UPCs at $\sqrt{s_{_{\rm NN}}}=5.02$ TeV.
For the sake of comparison, we also juxtapose the prediction by Ajjath et.al~\cite{AH:2023kor} with our prediction.
By integrating the joint dependencies of $\bm{q}_\perp$
 and $\bm{b}_\perp$, our predictions modestly enhance the cross-section magnitude compared to~\cite{AH:2023kor}, aligning more closely with the trend observed in the {\tt ATLAS} measurement.

\begin{figure}
\centering\includegraphics[scale=0.35]{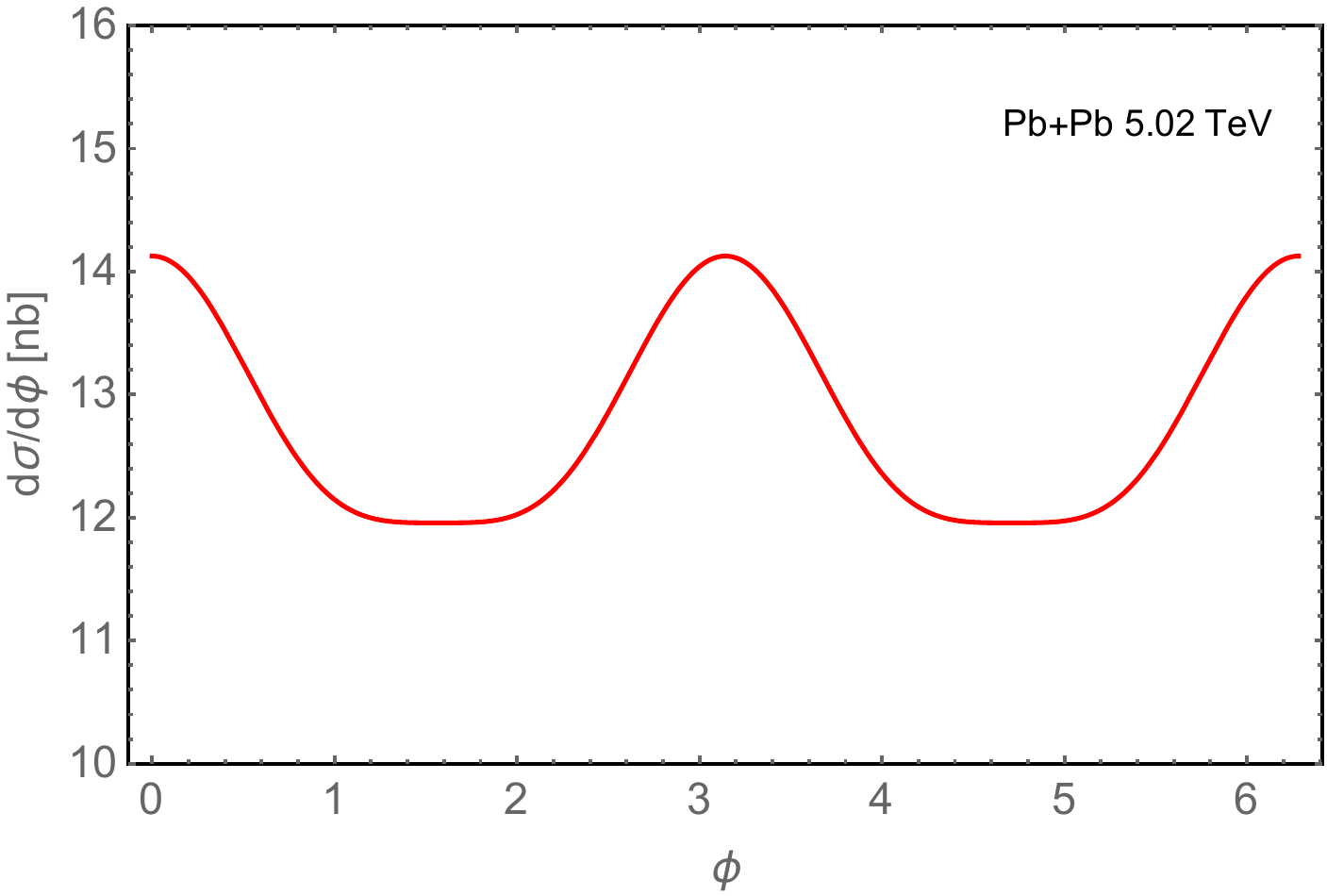}
\caption{The differential cross section of LbL scattering with respect to the azimuthal angle $\phi$ in Pb-Pb UPCs at $\sqrt{s_{_{\rm NN}}}=5.02$  TeV. 
The following cuts are imposed: $Q>5$ GeV, $P_\bot >$ 2.5 GeV,
and the rapidity of each outgoing photon satisfies $|y_{1,2}|<2.37$.
\label{fig:phi}}
\end{figure}

\begin{figure}
\centering\includegraphics[scale=0.35]{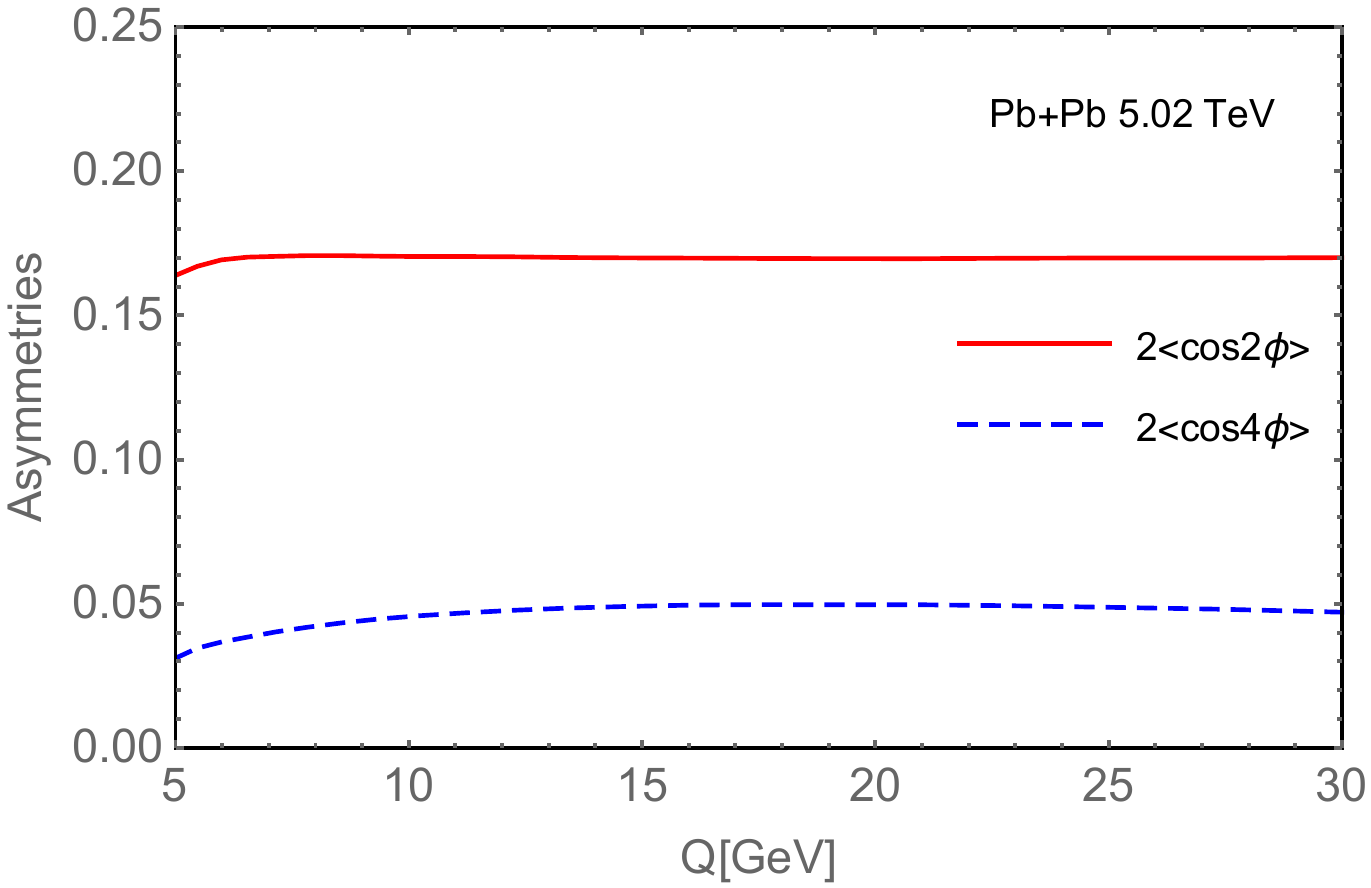}
\centering\includegraphics[scale=0.35]{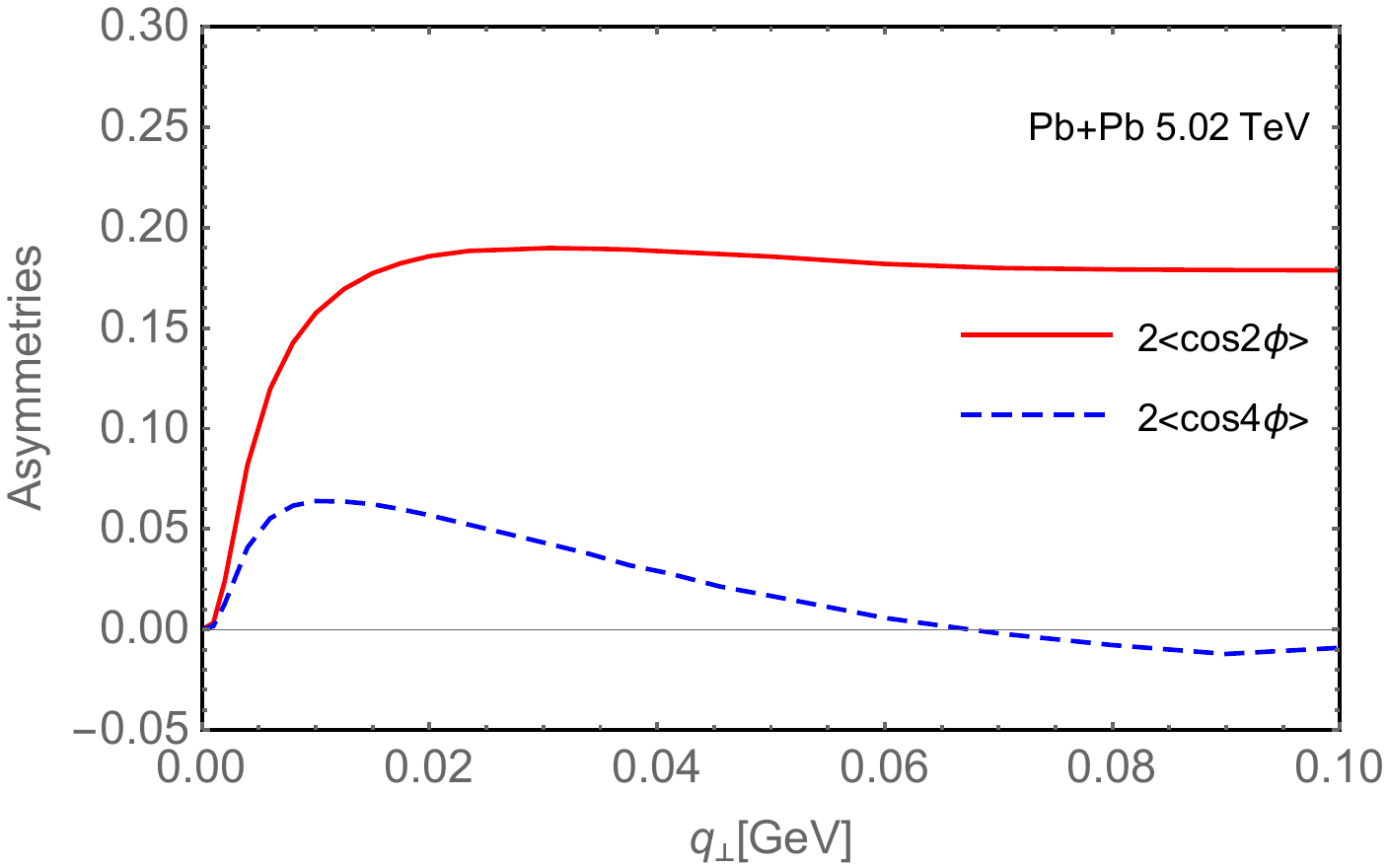}
\caption{The azimuthal asymmetries of LbL scattering with respect to the invariant mass Q (upper panel, with the total transverse
momentum of the diphoton, $q_\bot$, to be less than 1 GeV),  and with respect to $q_\bot$ (lower panel, with 5 GeV $< Q <$ 30 GeV), in Pb-Pb UPCs at $\sqrt{s_{_{\rm NN}}}=5.02$ TeV. 
The cuts $P_\bot >$ 2.5 GeV and $|y_{1,2}|<2.37$ applies to both figures. 
\label{fig:cos}}
\end{figure}


Finally, we show the differential cross section with respect to the azimuthal angle $\phi$ in Fig.~\ref{fig:phi}, from which
one clearly observes the sizable azimuthal modulation of $\cos 2\phi$ type.
Similar azimuthal modulation pattern has been observed in BW process and diffractive vector meson production in UPCs~\cite{Li:2019sin,Li:2019yzy,Xing:2020hwh}.
As mentioned before, all these phenomena have a common physical root:
the linear polarization of coherent photons induced by relativistic heavy ions.

In order to quantify the magnitude of the azimuthal asymmetry, we introduce the variable $\langle \cos 2\phi \rangle$ by
\beq
\left\langle \cos(2\phi)\right\rangle \equiv\frac{\int d\sigma\cos(2\phi)}{\int d\sigma}.
\eeq

The upper panel of Fig.~\ref{fig:cos}  investigates the dependence of the azimuthal asymmetry on the photon pair invariant mass $Q$, spanning a broader range from 5 to 30 GeV. Here, both $2\langle \cos(2\phi)$
 and $2\langle \cos(4\phi)$
 are plotted, revealing that the $\cos 2\phi$
 asymmetry remains relatively stable, fluctuating around 16\% across the entire $Q$ range, whereas the $\cos 4\phi$
 asymmetry is consistently lower, remaining  below 5\% for$Q > 10$ GeV.

As illustrated in the lower pannel of Fig.~\ref{fig:cos}, the asymmetry $2\left\langle \cos(2\phi)\right\rangle $
 reaches approximately 17\% at $q_\perp \sim 0.02$ GeV, and then remains relatively stable when $q_\perp> 0.02$ GeV. The lower panel also shows the behavior of $2\langle \cos(4\phi)\rangle$, which is smaller in magnitude, peaking at around 5\% and decreasing more rapidly as the photon transverse momentum $q_\perp$
 increases beyond 0.04 GeV. 
 The predicted asymmetries are within the reach of future measurements at the LHC, offering a promising opportunity to validate these theoretical insights experimentally. Additionally, we have computed the azimuthal dependent  cross section for LbL scattering in eA collisions at EIC energies, with the results detailed in the appendix.

\vspace{0.2cm}
\noindent{\color{blue}\it Summary---}
In this work we have initiated the study of linearly polarized light by linearly polarized light scattering in high energy collisions. 
The linear polarization of the initial state photons manifests in the non-trivial azimuthal distribution of the final state scattered photons. Specifically, we compute the azimuthal-dependent differential cross section of LbL scattering in UPCs  and find that the linear polarization of photons coherently emitted from 
relativistic heavy ions induces a significant $\cos 2\phi$ azimuthal modulation.
This discovery may open a new avenue for probing photon interactions with the QED vacuum 
under extreme conditions. It is also worth mentioning that the LbL scattering process has served a sensitive probe to search for the footprint of various BSM scenarios.
It is conceivable that the azimuthal modulation, as a novel observable supplementary to the familiar observables
such as diphoton invariant mass and rapidity, may be of great utility to discriminate between different BSM models.

\begin{acknowledgments}
We are grateful to Ji-chen Pan for participating in the early stage of this work, and Siwei Hu for useful discussion.
We also thank  Hua-sheng Shao and Xiao-nu Xiong for helpful discussions.
The work of Y.~J. is supported in part by the National Science Foundations of China under Grant No.~11925506.
The work of J. Z. is supported in part by the National Science Foundations of China under Grant No.~12175118, and No.~12321005.
The work of Y. Z. is supported in part by the Natural Science Foundation of China under Grant No.~12475084, and Shandong Province Natural Science Foundation under Grant No. ZR2024MA012.
\end{acknowledgments}

\appendix

\section{LbL scattering in $eA$ collisions}

It is also interesting to explore the possibility of observing the LbL scattering in the future electron-ion collision facilities, exemplified by 
{\tt EIC}~\cite{AbdulKhalek:2021gbh}.
In this appendix, we investigate the azimuthal modulation in the LbL scattering process, which is induced by the linearly polarized photons 
radiated off both from the electron and heavy ion.  

For definiteness, let us take the benchmark energy at {\tt EIC} to be $E_e=18$ GeV and $E_A=100$ GeV, respectively.
Some important simplification can be made in the $e+$Au collision relative to the Pb+Pb UPC. 
Since the electron is a point-like particle, it is legitimate to integrate \eqref{eq:cross section} 
over the impact parameter from 0 to $\infty$. 
We end up with the following differential cross section for LbL scattering in $eA$ collision in the laboratory frame:
\begin{eqnarray}
 &  & \frac{d\sigma}{d^{2}\bm{p}_{1\perp}d^{2}\bm{p}_{2\perp}dy_{1}dy_{2}}\!
 \nn \\
 & = & \!\frac{1}{32\pi^{2}Q^{4}}\int d^{2}\bm{k}_{1\perp}d^{2}\bm{k}_{2\perp}\delta^{2}(q_{\perp}-\bm{k}_{1\perp}-\bm{k}_{2\perp})x_{1}x_{2}
\nn \\
 & \times  & \bigg\{ \frac{1}{2}\left(|M_{+-}|^{2}+|M_{++}|^{2}\right)f_{1e}(x_{1},\bm{k}_{1\perp}^{2})f_1(x_{2},\bm{k}_{2\perp}^{2})
 \nn \\
 & - & \cos2(\phi_{1}){\text{R}e}[M_{++}M_{+-}^{*}]h_{1e}^{\perp}(x_{1},\bm{k}_{1\perp}^{2})f_1(x_{2},\bm{k}_{2\perp}^{2})
 \nn \\
 & - & \cos2(\phi_{2}){\text{R}e}[M_{++}M_{+-}^{*}]f_{1e}(x_{1},\bm{k}_{1\perp}^{2})h_{1}^{\perp}(x_{2},\bm{k}_{2\perp}^{2})
 \nn \\
 & + & \frac{1}{2} \Big[\cos2(\phi_{1}-\phi_{2})|M_{++}|^{2}+\cos2(\phi_{1}+\phi_{2})|M_{+-}|^{2}\Big]
 \nn \\
 &  & \times h^{\perp}_{1e}(x_{1},\bm{k}_{1\perp}^{2})h_{1}^{\perp}(x_{2},\bm{k}_{2\perp}^{2})\bigg\}. 
 \label{eq:eic}
\eqa
where the subscript `$e$' is 
utilized to emphasize that the photon TMD is affiliated with the electron. 

The photon TMDs inside an electron can be computed in pertubation theory, whose LO expressions are
\begin{subequations}
\bqa
&  & f_{1e}(x,\bm{k}_{\bot}^{2}) = \frac{\alpha_{e}}{2\pi^{2}}\frac{1+(1-x)^{2}}{x}\frac{\bm{k}_{\bot}^{2}}{(\bm{k}_{\bot}^{2}+x^{2}m_{e}^{2})^{2}},
\\
&  &  h_{1e}^{\perp}(x,\bm{k}_{\bot}^{2}) = \frac{\alpha_{e}}{\pi^{2}}\frac{1-x}{x}\frac{\bm{k}_{\bot}^{2}}{(\bm{k}_{\bot}^{2}+x^{2}m_{e}^{2})^{2}}.
\eqa
\end{subequations}

\begin{figure}
\centering\includegraphics[scale=0.38]{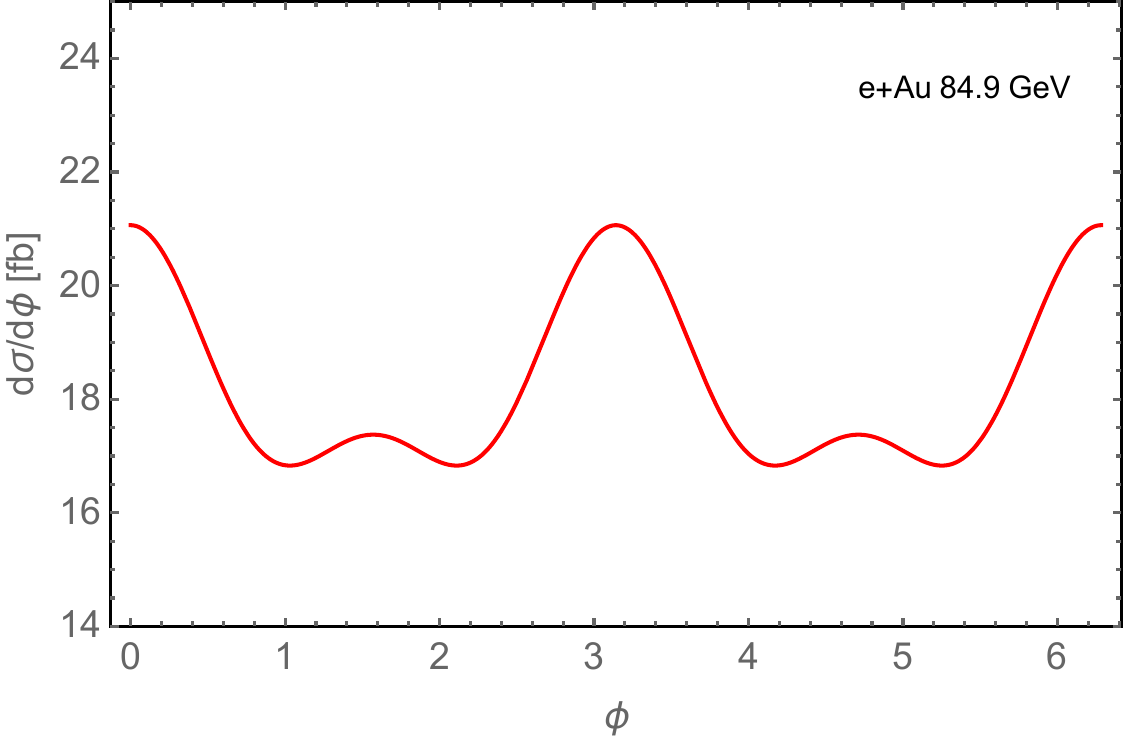}
\caption{The differential cross section of LbL scattering with respect to the azimuthal angle $\phi$ in an $e$+Au collision at EIC energy. We impose the following cuts: $P_\bot >$ 200 MeV, $\text {5 GeV} < Q <$ 15 GeV, $q_\bot < 100$ MeV, $-2<y_{1}<-1$ and $1<y_{2}<2$.
\label{fig:phi_EIC}}
\end{figure}

\begin{figure}
\centering\includegraphics[scale=0.38]{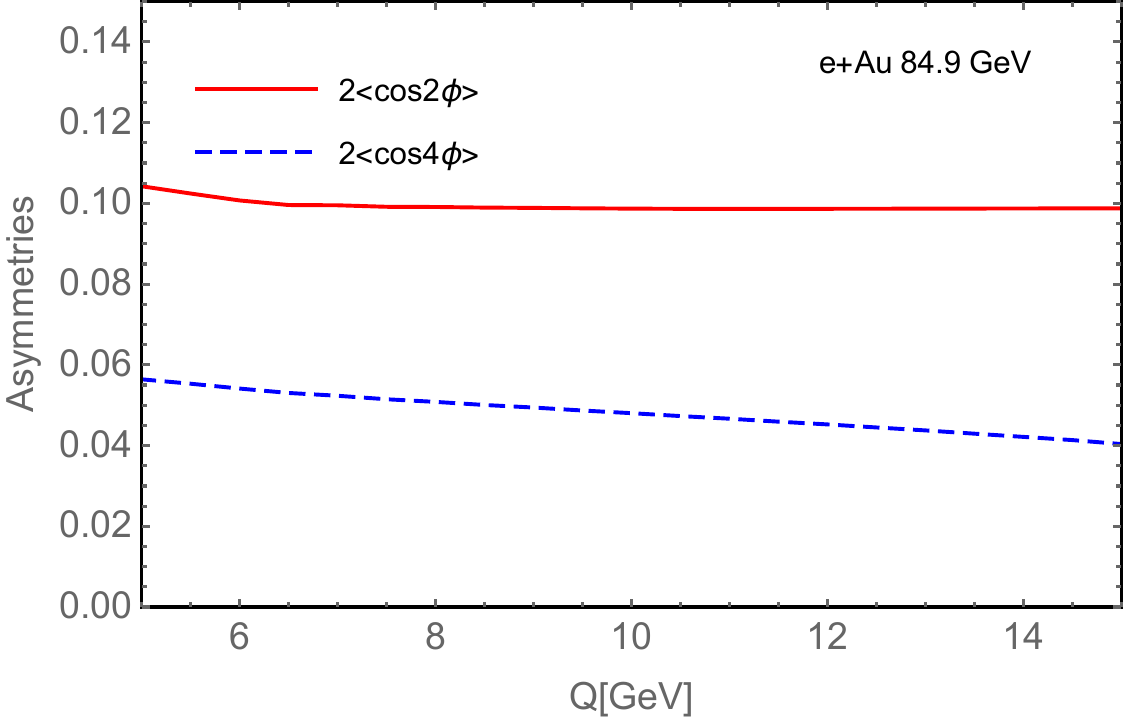}
\centering\includegraphics[scale=0.38]{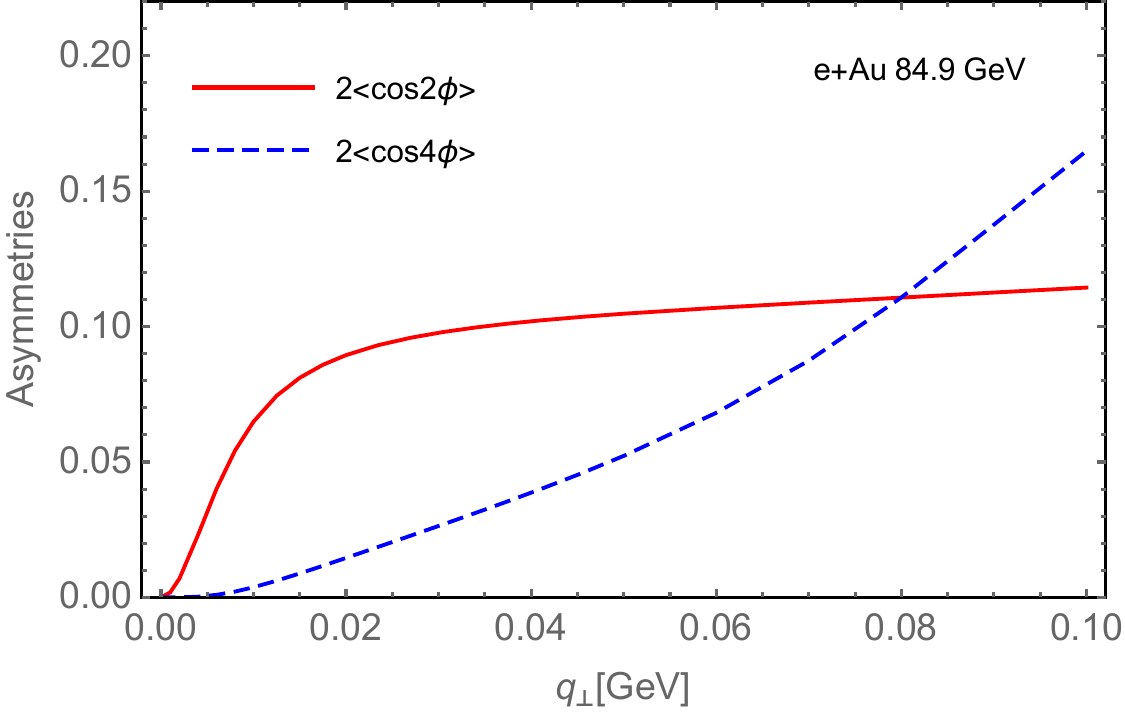}
\caption{The azimuthal asymmetries of LbL scattering with respect to the invariant mass Q (upper panel, with $q_\bot < 100$ MeV,  and with respect to $q_\bot$ (lower panel, with $\text {5 GeV} < Q <$ 15 GeV), in e-Au collisions at EIC energy. The cuts $P_\bot >$ 200 MeV, $-2<y_{1}<-1$ and $1<y_{2}<2$ applies to both figures. 
\label{fig:cos_EIC}}
\end{figure}




The photon TMDs of a heavy ion have been given in \eqref{eq:photonTMD}. The longitudinal momentum fractions of the incoming 
photons are connected with the rapidities of the outgoing photons through $x_{1}=\frac{P_{\perp}}{2E_{e}}(e^{ y_{1}}+e^{ y_{2}})$ and $x_{2}=\frac{P_{\perp}}{2E_{A}}(e^{- y_{1}}+e^{- y_{2}})$.

In the numerical analysis, we impose the following kinematic cuts. The rapidities of the outgoing photons are restricted in the range 
$-2<y_{1}<-1$, $1<y_{2}<2$, the total transverse momentum of the photon pair is less than $0.1$ GeV. We also require $P_{\perp}>0.2 ~\text{GeV}$ and Q > 5 GeV.
The total total cross section is predicted to be approximately 11 fb in the above kinematica region. Assuming the integrated luminosity about $10\;{\rm fb}^{-1}$ per year for $e$+Au run at {\tt EIC},
the number of LbL scattering events are expected to reach $110$.

In Fig.~\ref{fig:phi_EIC}, we plot the cross section  differential in the azimuthal angle $\phi$, and in In Fig.~\ref{fig:cos_EIC}, we plot the  
$\left\langle \cos2\phi\right\rangle $ as a function of $Q$ (upper panel) and $q_\perp$ (lower panel).
As in the UPC case, a pronounced $\cos2\phi$ azimuthal modulation is observed, which is a consequence of the linearly polarized incident photons. 
From Fig.~\ref{fig:cos_EIC}, we observe that the $2 \left\langle \cos2\phi\right\rangle$ 
 is approximately 10 \% .
 
The computed yields highlight a promising avenue for experimental investigations of  light-by-light scattering at future {\tt EIC} experiments, particularly in measuring azimuthal asymmetry.







\bibliography{ref}

\end{document}